\documentclass[english,pra]{revtex4}
\usepackage[T1]{fontenc}
\usepackage[latin9]{inputenc}
\usepackage{verbatim}
\usepackage{amsmath}
\usepackage{amssymb}
\usepackage{esint}

\makeatletter

\providecommand{\tabularnewline}{\\}

\usepackage{babel}

\makeatother

\usepackage{babel}
\begin{document}

\title{Weak measurement with orthogonal pre-selection and post-selection}

\author{Shengshi Pang}

\email{pangss@mail.ustc.edu.cn}

\author{Shengjun Wu}

\email{shengjun@ustc.edu.cn}

\author{Zeng-Bing Chen}

\email{zbchen@ustc.edu.cn}

\affiliation{Hefei National Laboratory for Physical Sciences at Microscale and
Department of Modern Physics, University of Science and Technology
of China, Hefei, Anhui 230026, China}

\begin{abstract}
Weak measurement is a novel quantum measurement scheme, which is usually
characterized by the weak value formalism. To guarantee the validity
of the weak value formalism, the fidelity between the pre-selection
and the post-selection should not be too small generally. In this
work, we study the weak measurement on a qubit system with exactly
or asymptotically orthogonal pre- and post-selections. We shall establish
a general rigorous framework for the weak measurement beyond the weak
value formalism, and obtain the average output of a weak measurement
when the pre- and post-selections are exactly orthogonal. We shall
also study the asymptotic behavior of a weak measurement in the limiting
process that the pre- and post-selections tend to be orthogonal.
\end{abstract}
\maketitle

\section{Introduction}

In the conventional quantum theory, a quantum measurement consists
of a set of probabilistic orthogonal projections onto the eigenstates
(or eigenspaces if any degeneracy) of an observable \cite{nielsen}.
The output of a quantum measurement ranges between the minimal eigenvalue
and the maximal eigenvalue of the observable. Such an ideal quantum
measurement can be realized if the spread of the wave function of
the probe is sufficiently sharp.

In 1988, Aharonov, Albert, and Vaidman (AAV) proposed a novel quantum
measurement scheme called \emph{weak measurement} \cite{aharonov}.
A weak measurement involves three stages generally: first, the system
to be measured and a measuring probe are prepared in initial states;
then the system and the probe are coupled by such a weak interaction
that the state of the system remains almost undisturbed after the
interaction; lastly, the system is observed, and if it is found to
be in a specific final state $|\psi_{f}\rangle$, the pointer shift
of the probe is recorded, otherwise, discarded. Usually, the initial
state of the system in the first stage is called pre-selection, and
the specific choice of the final state of the system in the last stage
is called post-selection. At the beginning, the new concept of weak
measurement aroused some controversy, but soon its physical significance
was clarified by \cite{prd}.

In contrast to the conventional orthogonal projective measurement,
the spread of the initial wave function of the probe is usually chosen
to be quite wide in a weak measurement in order that its output can
be far beyond the range of the eigenvalues of the observable on the
system. The magic large pointer shift of the probe in a weak measurement
is attributed to the interference in the superposition of several
slightly different probe states after the post-selection on the system.

Since the birth of weak measurement, a lot of research has been devoted
to this interesting field, including weak measurement with arbitrary
probe \cite{arbitrary probe}, weak measurement involving the contribution
of probe dynamics \cite{probe dynamics}, weak measurement with entangled
probes \cite{entangled probes}, weak measurement with a qubit probe
\cite{wu pla}, weak measurement with a probe in a mixed state \cite{incoherent probe},
geometric phase in weak measurement \cite{geometric phase pla,geometric phase njp},
continuous quantum measurement of coherent oscillations between two
quantum states of an individual two-state system \cite{coherent oscillation},
and so on. Moreover, weak measurement has been found universal to
implement any general quantum measurement \cite{universal}.

In recent years, weak measurement has been realized in experiment
\cite{experiment-weak value}. Besides, weak measurement has also
been used to experimentally examine quantum paradoxes \cite{paradoxes}
and for experimental feedback control of quantum systems in the presence
of noise \cite{feedback}. And more experiment protocols using weak
measurements have been proposed \cite{proposal-arrival2,proposal-arrivaltime,proposal-chargesensing,proposal-communication,proposal-phaseshift,proposal-wu-marek}.

An important physical quantity in weak measurement is\emph{ }the \emph{weak
value} introduced in \cite{aharonov}, and it is defined as
\begin{equation}
A_{w}=\frac{\langle\psi_{f}|A|\psi_{i}\rangle}{\langle\psi_{f}|\psi_{i}\rangle},\label{eq:0}
\end{equation}
where $A$ is an observable acting on the system, and $|\psi_{i}\rangle$,
$|\psi_{f}\rangle$ are the pre-selection and post-selection of the
system state respectively. Roughly speaking, weak value characterizes
the pointer shift of the probe after a weak measurement. Weak value
has also received intensive study \cite{weakvalue-joza,weakvalue1,weakvalue4,weakvalue5,weakvalue6,weakvalue8}.

It can be seen that when the fidelity between the pre- and post-selections
$\langle\psi_{f}|\psi_{i}\rangle$ is small in \eqref{eq:0}, the
weak value $A_{w}$ can be very large, indicating a large position
shift of the probe. This important characteristic of weak measurement
is found useful to amplify and measure tiny physical quantities which
are difficult to detect by conventional techniques in experiments.
For example, weak measurement has been used to observe the spin Hall
effect of light \cite{science spin hall}, optical beam deflection
\cite{signaltonoise,signaltonoise2} and optical frequency change
\cite{precision}, etc.

However, a necessary condition for the validity of the weak value
formalism is that the fidelity between the pre- and post-selections
should not be too small \cite{aharonov,prd}. Then, some interesting
questions arise naturally: what is the result of a weak measurement
when the fidelity between the pre- and post-selections is very small,
or even zero? And is it possible that a weak measurement has larger,
or even infinitely large amplification effect in such cases?

In the literature, it has been noticed that the original AAV\textquoteright{}s
weak value formalism breaks down in a weak measurement with nearly
orthogonal pre- and post-selections. Specific cases such as a Sagnac
interferometer with continuous phase amplification \cite{Sagnac}
and the original Stern-Gerlach setup \cite{stern galac} were studied
in such situations. In \cite{liyang}, some general results of weak
measurement have been derived, and orthogonal weak value was introduced
as well. In \cite{zhuxuanmin}, the maximum position shift of a Gaussian
pointer is obtained for a qubit system and the signal-to-noise ratio
is studied in detail.

The purpose of this article is to provide a more rigorous treatment
of the weak measurement for a qubit system interacting with a continuous
probe, when the pre- and post-selections are exactly or asymptotically
orthogonal. We shall derive rigorous general results without any approximation,
which can be readily applied to cases with specified pre- and post-selections,
and we shall also derive the asymptotic behavior of the output for
a weak measurement with pre- and post-selections approaching orthogonality
in various cases.

The structure of this paper is as follows. We shall first review the
weak value formalism in Sec. II, then establish a general framework
for weak measurement which is suitable for studying a weak measurement
with variable fidelity between the pre- and post-selections in Sec.
III. In that section, we shall also prove that the amplification effect
of a weak measurement cannot be infinitely large, and obtain the analytical
average output of a weak measurement when the pre- and post-selections
are exactly orthogonal. In Sec. IV, we shall discuss the case with
asymptotically orthogonal pre- and post-selections in detail.

\section{Preliminary}

In a weak measurement, a typical Hamiltonian in the interaction picture
is
\begin{equation}
H_{\mathrm{int}}=g(t)\hat{A}\otimes\hat{p},\; g(t)=g\delta(t-t_{0}),\label{eq:1}
\end{equation}
 where $g$ is a small coupling constant and $\hat{p}$ is the momentum
operator on the probe conjugate to the position operator $\hat{q}$.
The time factor $\delta(t-t_{0})$ means that the weak interaction
happens only for a very short instant.

Suppose the initial states of the system and the probe are $|\psi_{i}\rangle$
and $|\phi\rangle$ respectively. When the weak interaction is finished,
the total state of the system and the probe evolves to
\begin{equation}
e^{-ig\hat{A}\otimes\hat{p}}|\psi_{i}\rangle\otimes|\phi\rangle.\label{eq:2}
\end{equation}
When the post-selection on the system is $|\psi_{f}\rangle$, the
state of the probe collapses to
\begin{equation}
\frac{\langle\psi_{f}|e^{-ig\hat{A}\otimes\hat{p}}|\psi_{i}\rangle\otimes|\phi\rangle}{|\langle\psi_{f}|e^{-ig\hat{A}\otimes\hat{p}}|\psi_{i}\rangle|}.\label{eq:3}
\end{equation}

When $g$ is sufficiently small, the position shift of the probe is
roughly proportional to the weak value $A_{w}$ defined in \eqref{eq:0}.
It should be noted that in \cite{aharonov} the Hamiltonian was $g(t)\hat{A}\otimes\hat{q}$
and the weak value actually characterized the momentum shift of the
probe.

The weak value $A_{w}$ can often be complex, and it has a beautiful
physical interpretation given by \cite{weakvalue-joza}: the real
part and imaginary part of the weak value are responsible for the
position shift and momentum shift of the probe state induced by the
weak measurement respectively. Specifically, it was shown in \cite{weakvalue-joza}
that
\begin{equation}
\begin{aligned}\langle\hat{q}\rangle_{f}= & \langle\hat{q}\rangle_{i}+g\mathrm{Re}A_{w}+g\mathrm{Im}A_{w}(m\frac{\mathrm{d}}{\mathrm{d}t}\mathrm{Var}_{\hat{q}}),\\
\langle\hat{p}\rangle_{f}= & \langle\hat{p}\rangle_{i}+2g\mathrm{Im}A_{w}\mathrm{Var}_{\hat{p}},
\end{aligned}
\label{eq:4}
\end{equation}
where $\mathrm{Var}_{\hat{q}}$, $\mathrm{Var}_{\hat{p}}$ are the
square deviations of position and momentum of the initial probe state
$|\phi\rangle$.

From Eq. \eqref{eq:4}, it can be seen that when the fidelity between
the pre- and post-selections is small, the position shift and momentum
shift of the probe state can be very large. And this phenomenon has
been used to magnify and observe small physical quantities in experiments,
as referred to in the last section.

However, the weak value formalism could not be applied to the situation
$\langle\psi_{f}|\psi_{i}\rangle\rightarrow0$ because $A_{w}\rightarrow\infty$
(and thus $\langle\hat{q}\rangle_{f},\,\langle\hat{p}\rangle_{f}\rightarrow\infty$)
in this situation. Such a divergence is obviously non-physical, and
it results from the derivation of the weak value, in which only the
terms of $g$ up to the first order were considered in the expansion
of the Hamiltonian \eqref{eq:1} \cite{aharonov}. That approximation
would be no longer valid if $\langle\psi_{f}|\psi_{i}\rangle\rightarrow0$,
therefore, in order to study the average output from the probe in
a weak measurement when the pre- and post-selections are exactly or
nearly orthogonal, such an approximation should be avoided.

In the following sections, we shall establish a new rigorous framework
for a general weak measurement on a qubit system with a continuous
probe beyond the weak value formalism, and derive the results of the
weak measurement when $\langle\psi_{f}|\psi_{i}\rangle=0$ or $\langle\psi_{f}|\psi_{i}\rangle\rightarrow0$.

\section{Weak measurement with exactly orthogonal pre- and post-selections}

\subsection{General result}

In this section, we shall first derive a general exact formula for
the average pointer reading of the probe after the post-selection,
then obtain an analytical result for the situation that the pre- and
post-selections are exactly orthogonal. In particular, the case that
the observable on the probe is the position operator $\hat{q}$ or
the momentum operator $\hat{p}$ will be studied respectively in detail.
The dimension of the system will be assumed to be two throughout this
paper.

Suppose the initial state of the probe is $|\phi\rangle$, and the
post-selection and pre-selection satisfy
\begin{equation}
|\psi_{f}\rangle=\alpha|\psi_{i}\rangle+\sqrt{1-\alpha^{2}}|\psi_{i}^{\perp}\rangle,\;\alpha\geq0.\label{eq:6}
\end{equation}
 Then, the fidelity between the post-selection and pre-selection is
\begin{equation}
|\langle\psi_{f}|\psi_{i}\rangle|=\alpha.\label{eq:7}
\end{equation}

The total state of the system and the probe after the unitary evolution
is
\begin{equation}
e^{-ig\hat{A}\otimes\hat{p}}|\psi_{i}\rangle\otimes|\phi\rangle,\label{eq:8}
\end{equation}
 so the probe state after the post-selection is
\begin{equation}
\frac{\langle\psi_{f}|e^{-ig\hat{A}\otimes\hat{p}}|\psi_{i}\rangle\otimes|\phi\rangle}{|\langle\psi_{f}|e^{-ig\hat{A}\otimes\hat{p}}|\psi_{i}\rangle\otimes|\phi\rangle|}.\label{eq:9}
\end{equation}
 Suppose the observable on the probe to be observed after the post-selection
is $\hat{M}$, then the average reading from the probe is
\begin{equation}
\langle\hat{M}\rangle=\frac{{\displaystyle \langle\psi_{i}|\otimes\langle\phi|e^{ig\hat{A}\otimes\hat{p}}|\psi_{f}\rangle}M{\displaystyle \langle\psi_{f}|e^{-ig\hat{A}\otimes\hat{p}}|\psi_{i}\rangle\otimes|\phi\rangle}}{{\displaystyle \langle\psi_{i}|\otimes\langle\phi|e^{ig\hat{A}\otimes\hat{p}}|\psi_{f}\rangle}{\displaystyle \langle\psi_{f}|e^{-ig\hat{A}\otimes\hat{p}}|\psi_{i}\rangle\otimes|\phi\rangle}}.\label{eq:10}
\end{equation}

Since the system is two-dimensional, the operator $\hat{A}$ has a
spectral decomposition
\begin{equation}
\hat{A}=a_{1}|a_{1}\rangle\langle a_{1}|+a_{2}|a_{2}\rangle\langle a_{2}|,\label{eq:11}
\end{equation}
 and
\begin{equation}
e^{-ig\hat{A}\otimes\hat{p}}=|a_{1}\rangle\langle a_{1}|e^{-iga_{1}\hat{p}}+|a_{2}\rangle\langle a_{2}|e^{-iga_{2}\hat{p}}.\label{eq:12}
\end{equation}
 Thus
\begin{equation}
\begin{aligned}\langle\psi_{f}|e^{-ig\hat{A}\otimes\hat{p}}|\psi_{i}\rangle & =\langle\psi_{f}|a_{1}\rangle\langle a_{1}|\psi_{i}\rangle e^{-iga_{1}\hat{p}}+\langle\psi_{f}|a_{2}\rangle\langle a_{2}|\psi_{i}\rangle e^{-iga_{2}\hat{p}}\\
 & =\alpha(|\langle a_{1}|\psi_{i}\rangle|^{2}e^{-iga_{1}\hat{p}}+|\langle a_{2}|\psi_{i}\rangle|^{2}e^{-iga_{2}\hat{p}})\\
 & +\sqrt{1-\alpha^{2}}(\langle\psi_{i}^{\perp}|a_{1}\rangle\langle a_{1}|\psi_{i}\rangle e^{-iga_{1}\hat{p}}+\langle\psi_{i}^{\perp}|a_{2}\rangle\langle a_{2}|\psi_{i}\rangle e^{-iga_{2}\hat{p}}).
\end{aligned}
\label{eq:5}
\end{equation}

As the phases of $|a_{1}\rangle$ and $|a_{2}\rangle$ can be arbitrary,
we can assume that $\langle a_{1}|\psi_{i}\rangle\geq0,\,\langle a_{2}|\psi_{i}\rangle\geq0$,
and
\begin{equation}
\begin{aligned}\langle a_{1}|\psi_{i}\rangle & =x, & \langle a_{2}|\psi_{i}\rangle & =\sqrt{1-x^{2}},\\
\langle a_{1}|\psi_{i}^{\perp}\rangle & =\sqrt{1-x^{2}}e^{i\theta}, & \langle a_{2}|\psi_{i}^{\perp}\rangle & =xe^{i\theta^{\prime}},
\end{aligned}
\; x\geq0.\label{eq:13}
\end{equation}

Note that
\begin{equation}
\begin{aligned}\langle\psi_{i}|\psi_{i}^{\perp}\rangle & =\langle\psi_{i}|(|a_{1}\rangle\langle a_{1}|+|a_{2}\rangle\langle a_{2}|)|\psi_{i}^{\perp}\rangle\\
 & =x\sqrt{1-x^{2}}(e^{i\theta}+e^{i\theta^{\prime}})\\
 & =0,
\end{aligned}
\label{eq:17}
\end{equation}
 so
\begin{equation}
\theta-\theta^{\prime}=\pi.\label{eq:18}
\end{equation}

Plugging Eq. \eqref{eq:13} into \eqref{eq:5}, we have
\begin{equation}
\langle\psi_{f}|e^{-ig\hat{A}\otimes\hat{p}}|\psi_{i}\rangle=\alpha(x^{2}e^{-iga_{1}\hat{p}}+(1-x^{2})e^{-iga_{2}\hat{p}})+\sqrt{1-\alpha^{2}}x\sqrt{1-x^{2}}e^{-i\theta}(e^{-iga_{1}\hat{p}}-e^{-iga_{2}\hat{p}}).\label{eq:14}
\end{equation}

Now, we define some notations for convenience. Let
\begin{equation}
\begin{aligned}W_{ij} & =\langle\phi|e^{iga_{i}\hat{p}}\hat{M}e^{-iga_{j}\hat{p}}|\phi\rangle=\int\phi^{*}(q-ga_{i})\hat{M}\phi(q-ga_{j})dq,\\
Y_{ij} & =\langle\phi|e^{ig(a_{i}-a_{j})\hat{p}}|\phi\rangle=\int\phi^{*}(q-ga_{i})\phi(q-ga_{j})dq.
\end{aligned}
\label{eq:15}
\end{equation}

Then, by plugging Eq. \eqref{eq:14} into \eqref{eq:10} with the
notations \eqref{eq:15}, we can get the expectation value of the
observable $M$ on the probe after post-selection. %
To simplify the calculation, let us define
\begin{equation}
\beta=\frac{\alpha}{\sqrt{1-\alpha^{2}}},\: y=\frac{x}{\sqrt{1-x^{2}}},\label{eq:22}
\end{equation}
 then it can be worked out that
\begin{equation}
\langle\hat{M}\rangle=\frac{\begin{aligned}\beta^{2}(y^{4}W_{11}+W_{22}+2y^{2}\mathrm{Re}W_{12})+y^{2}(W_{11}+W_{22}-2\mathrm{Re}W_{12})\\
+2\beta y[y^{2}((W_{11}-\mathrm{Re}W_{12})\cos\theta-\mathrm{Im}W_{12}\sin\theta)+((\mathrm{Re}W_{12}-W_{22})\cos\theta-\mathrm{Im}W_{12}\sin\theta)]
\end{aligned}
}{\begin{aligned}\beta^{2}(y^{4}+1+2y^{2}\mathrm{Re}Y_{12})+2y^{2}(1-\mathrm{Re}Y_{12})+2\beta y[(y^{2}-1)(1-\mathrm{Re}Y_{12})\cos\theta-(y^{2}+1)\mathrm{Im}Y_{12}\sin\theta]\end{aligned}
}.\label{eq:23}
\end{equation}

Eq. \eqref{eq:23} is a general exact formula for the expectation
value of an observable $\hat{M}$ on the probe after a weak measurement,
and it is the starting point of our further study on the asymptotic
property of $\langle\hat{M}\rangle$ when $\langle\psi_{f}|\psi_{i}\rangle\rightarrow0$.

In particular, when $|\langle\psi_{f}|\psi_{i}\rangle|=0$, i.e. $\beta=0$,
we can immediately obtain
\begin{equation}
\langle\hat{M}\rangle=\frac{W_{11}+W_{22}-2\mathrm{Re}W_{12}}{2(1-\mathrm{Re}Y_{12})}\label{eq:26}
\end{equation}
from Eq. \eqref{eq:23}.

From \eqref{eq:15}, $\mathrm{Re}Y_{12}<1$, so $\langle\hat{M}\rangle$
can never be infinity when $|\langle\psi_{f}|\psi_{i}\rangle|=0$,
in sharp contrast to \eqref{eq:0} and \eqref{eq:4} which indicate
that $\langle\hat{M}\rangle$ diverges if $|\langle\psi_{f}|\psi_{i}\rangle|=0$.

And it can also be proved that the amplification effect of a weak
measurement cannot be infinitely large when $|\langle\psi_{f}|\psi_{i}\rangle|\neq0$.
From Eq. \eqref{eq:23}, it can be seen that $\langle\hat{M}\rangle$
can be infinity only when the denominator is zero. Since the denominator
of \eqref{eq:23} is a quadratic polynomial of $\beta$, the necessary
condition that the denominator can be zero is that its discriminant
is non-negative. However, by some calculation, it turns out that
\begin{equation}
\begin{aligned}\Delta & =4y[(y-1)(1-\mathrm{Re}Y_{12})\cos\theta-(y+1)\mathrm{Im}Y_{12}\sin\theta]^{2}-8y(1-\mathrm{Re}Y_{12})(y^{2}+1+2y\mathrm{Re}Y_{12})\\
 & \leq4y[(y-1)^{2}(1-\mathrm{Re}Y_{12})^{2}+(y+1)^{2}(\mathrm{Im}Y_{12})^{2}]-8y(1-\mathrm{Re}Y_{12})(y^{2}+1+2y\mathrm{Re}Y_{12})\\
 & =-4y(y+1)^{2}(1-|Y_{12}|^{2})<0,
\end{aligned}
\label{eq:24}
\end{equation}
 so the denominator of \eqref{eq:23} cannot be zero and $\langle\hat{M}\rangle$
can never be infinity then.

An interesting thing to note in \eqref{eq:26} is that $\langle\hat{M}\rangle$
only depends on the initial state of the probe $|\phi\rangle$ and
the coupling constant $g$ when the pre- and post-selections are orthogonal.

\subsection{Examples}

In this subsection, we give some examples to illustrate the results
obtained in the last subsection. We shall calculate the expectation
value of the position and momentum of the probe after post-selection
respectively, in particular when the pre- and post-selections are
orthogonal.

\subsubsection{$\hat{M}=\hat{q}$}

If $\hat{M}=\hat{q}$, it can be easily verified that
\begin{equation}
\begin{aligned}W_{11} & =\langle\hat{q}\rangle_{i}+ga_{1},\\
W_{22} & =\langle\hat{q}\rangle_{i}+ga_{2},\\
W_{12} & =\int\phi^{*}(q-ga_{1})q\phi(q-ga_{2})dq,
\end{aligned}
\label{eq:27}
\end{equation}
where $\langle\hat{q}\rangle_{i}=\int\phi^{*}(q)q\phi(q)dq$ is the
expectation value of the probe position before the weak interaction.

Plugging the above equations into \eqref{eq:23}, we can get the mean
position shift of the probe after a weak measurement.

Particularly, when $|\langle\psi_{f}|\psi_{i}\rangle|=0$, i.e. $\alpha=0$,
\begin{equation}
\begin{aligned}\langle\hat{q}\rangle & =\frac{\langle\hat{q}\rangle_{i}+g(a_{1}+a_{2})/2-\mathrm{Re}W_{12}}{1-\mathrm{Re}Y_{12}}.\end{aligned}
\label{eq:25}
\end{equation}

Provided that $q\phi^{(n)}(q)\rightarrow0\,(n=0,1,2\cdots)$ as $q\rightarrow\pm\infty$
(this can be satisfied by most wave functions like the Gaussian-type
wave functions, but not by some wave functions which oscillate extremely
quickly as $q\rightarrow\pm\infty$), it can be worked out that
\begin{equation}
\begin{aligned}\mathrm{Re}Y_{12} & =\sum_{n=0}^{+\infty}\frac{(-1)^{n}}{(2n)!}g^{2n}(a_{1}-a_{2})^{2n}\int|\phi^{(n)}(q)|^{2}dq,\\
\mathrm{Re}W_{12} & =\sum_{n=0}^{+\infty}\frac{(-1)^{n}}{(2n)!}g^{2n}(a_{1}-a_{2})^{2n}\int q|\phi^{(n)}(q)|^{2}dq+\frac{1}{2}\sum_{n=0}^{+\infty}\frac{(-1)^{n}}{(2n)!}g^{2n+1}(a_{1}+a_{2})(a_{1}-a_{2})^{2n}\int|\phi^{(n)}(q)|^{2}dq.
\end{aligned}
\label{eq:68}
\end{equation}
The details of calculation are presented in the appendix.

Plugging Eq. \eqref{eq:68} into \eqref{eq:25}, we get
\begin{equation}
\begin{aligned}\langle\hat{q}\rangle & =\frac{1}{2}g(a_{1}+a_{2})+\frac{{\displaystyle \sum_{n=0}^{+\infty}\frac{(-1)^{n}}{(2n+2)!}g^{2n}(a_{1}-a_{2})^{2n}\int q|\phi^{(n+1)}(q)|^{2}dq}}{{\displaystyle \sum_{n=0}^{+\infty}\frac{(-1)^{n}}{(2n+2)!}g^{2n}(a_{1}-a_{2})^{2n}\int|\phi^{(n+1)}(q)|^{2}dq}}.\end{aligned}
\label{eq:49}
\end{equation}

A straightforward conclusion from Eq. \eqref{eq:49} is that if the
initial wave function $\phi(q)$ of the probe is symmetric or anti-symmetric,
\begin{equation}
\langle\hat{q}\rangle=\frac{1}{2}g(a_{1}+a_{2})\label{eq:50}
\end{equation}
when $|\langle\psi_{f}|\psi_{i}\rangle|=0$. This has been verified
by the case of $\phi(q)$ being an Gaussian state \cite{zhuxuanmin}.

In addition, when $g$ is small, only low orders of $g$ need to be
remained in Eq. \eqref{eq:49}, which can simplify the calculation
of $\langle\hat{q}\rangle$.

\subsubsection{$\hat{M}=\hat{p}$}

If $\hat{M}=\hat{p}$, then
\begin{equation}
\begin{aligned}W_{11} & =W_{22}=\langle\hat{p}\rangle_{i}=2\int\mathrm{Re}\phi(q)\mathrm{Im}\dot{\phi}(q)dq;\\
W_{12} & =\int\phi^{*}(q-ga_{1})\hat{p}\phi(q-ga_{2})dq=-i\int\phi^{*}(q-ga_{1})\dot{\phi}(q-ga_{2})dq.
\end{aligned}
\label{eq:28}
\end{equation}

By plugging \eqref{eq:28} into \eqref{eq:23}, one can get the mean
momentum shift of the probe after a weak measurement.

When the pre- and post-selections are orthogonal, i.e. $\alpha=0$,
\begin{equation}
\langle\hat{p}\rangle=\frac{\begin{aligned}\langle\hat{p}\rangle_{i}-\mathrm{Re}W_{12}\end{aligned}
}{\begin{aligned}1-\mathrm{Re}Y_{12}\end{aligned}
}.\label{eq:34}
\end{equation}

Provided that $q\phi^{(n)}(q)\rightarrow0\,(n=0,1,2\cdots)$ as $q\rightarrow\pm\infty$,
detailed calculation shows
\begin{equation}
\mathrm{Re}W_{12}=2\sum_{n=0}^{+\infty}\frac{(-1)^{n}g^{2n}}{(2n)!}(a_{1}-a_{2})^{2n}\int(\mathrm{Re}\phi(q))^{(n)}(\mathrm{Im}\phi(q))^{(n+1)}dq.\label{eq:69}
\end{equation}
The details of calculation are given in the appendix.

Therefore, when $|\langle\psi_{f}|\psi_{i}\rangle|=0$,
\begin{equation}
\langle\hat{p}\rangle=\frac{{\displaystyle \sum_{n=0}^{+\infty}\frac{(-1)^{n}}{(2n+2)!}g^{2n}(a_{1}-a_{2})^{2n}\int(\mathrm{Re}\phi(q))^{(n+1)}(\mathrm{Im}\phi(q))^{(n+2)}dq}}{{\displaystyle \sum_{n=0}^{+\infty}\frac{(-1)^{n}}{(2n+2)!}g^{2n}(a_{1}-a_{2})^{2n}\int|\phi^{(n+1)}(q)|^{2}dq}}.\label{eq:38}
\end{equation}

If the initial wave function $\phi(q)$ of the probe is symmetric
or anti-symmetric, it can be inferred from Eq. \eqref{eq:38} that
\begin{equation}
\langle\hat{p}\rangle=0\label{eq:47}
\end{equation}
 when $|\langle\psi_{f}|\psi_{i}\rangle|=0$.

\section{Weak measurement with asymptotically orthogonal pre- and post-selections}

In the last section, we gave a general rigorous framework for the
weak measurement on a qubit system with a continuous probe, and applied
it to the case of exactly orthogonal pre- and post-selections. In
this section, we study the weak measurement with asymptotically orthogonal
pre- and post-selections.

It is known that, when the pre- and post-selections are not extremely
orthogonal, the average output of a weak measurement $\langle\hat{M}\rangle$
can be characterized by the weak value \eqref{eq:0} \cite{aharonov,weakvalue-joza},
and it is roughly proportional to the reciprocal of the fidelity between
the pre- and post-selections. However, when the pre- and post-selections
tend to be orthogonal, the weak value would be no longer valid because
it may diverge. In the last section, it was shown that when the pre-
and post-selections are exactly orthogonal, the average output of
a weak measurement $\langle\hat{M}\rangle$ is still finite. That
gives a hint that when the pre- and post-selections tend to be orthogonal,
$\langle\hat{M}\rangle$ does not vary as the reciprocal of the fidelity
between them actually. So, it is interesting to study how the weak
measurement behaves in the limiting process that the pre- and post-selections
tend to be orthogonal, and this is what we shall focus on in this
section.

Let $|\psi_{i}\rangle$ and $|\psi_{f}\rangle$ denote the pre- selection
and the post-selection of the system as before. Note that in a limiting
process $\langle\psi_{f}|\psi_{i}\rangle\rightarrow0$, the fidelity
between the pre-selection and one of the eigenstates of the observable
$A$ on the system, either $x_{1}$ or $x_{2}$, can also vary. If
$x_{1},x_{2}\nrightarrow0\,\mathrm{or}\,1$, then $\langle\psi_{f}|\psi_{i}\rangle$
dominates the asymptotic behavior of $\langle\hat{M}\rangle$. However,
if $x\rightarrow0\,\mathrm{or}\,1$ as $\langle\psi_{f}|\psi_{i}\rangle\rightarrow0$,
i.e. the pre-selection tends towards one of the eigenstates of the
observable $A$, the problem becomes more complex, because $x_{1}$
or $x_{2}$ may give considerable contribution to $\langle\hat{M}\rangle$
and there will be competition between $\langle\psi_{f}|\psi_{i}\rangle$
and $x_{1}$ or $x_{2}$. So, whether $x_{1},x_{2}\rightarrow0\,\mathrm{or}\,1$
needs be taking into account in studying the asymptotic behavior of
$\langle\hat{M}\rangle$.

We first consider the relatively simpler case that $|\psi_{i}\rangle\nrightarrow|a_{1}\rangle\,\mathrm{or}\,|a_{2}\rangle$,
i.e. $x_{1},x_{2}\nrightarrow0\,\mathrm{or}\,1$ as $\langle\psi_{f}|\psi_{i}\rangle\rightarrow0$.
In this case, there is no competition between $\langle\psi_{f}|\psi_{i}\rangle$
and $x_{1}$ or $x_{2}$, and only $\langle\psi_{f}|\psi_{i}\rangle$
dominates $\langle\hat{M}\rangle$ when $\langle\psi_{f}|\psi_{i}\rangle\rightarrow0$.
Thus, the quadratic and higher order terms of $\beta$ can be omitted
in \eqref{eq:23}, and \eqref{eq:23} can be simplified to

\begin{equation}
\langle\hat{M}\rangle\approx\frac{y(W_{11}+W_{22}-2\mathrm{Re}W_{12})/2+\beta[y^{2}((W_{11}-\mathrm{Re}W_{12})\cos\theta-\mathrm{Im}W_{12}\sin\theta)+(\mathrm{Re}W_{12}-W_{22})\cos\theta-\mathrm{Im}W_{12}\sin\theta]}{y(1-\mathrm{Re}Y_{12})+\beta[(y^{2}-1)(1-\mathrm{Re}Y_{12})\cos\theta-(y^{2}+1)\mathrm{Im}Y_{12}\sin\theta]}.\label{eq:48}
\end{equation}

Considering
\begin{equation}
\frac{a_{1}z+b_{1}}{a_{2}z+b_{2}}\approx\frac{b_{1}}{b_{2}}+(\frac{a_{1}}{b_{2}}-\frac{a_{2}b_{1}}{b_{2}^{2}})z,\, z\rightarrow0,\label{eq:54}
\end{equation}
 we have
\begin{equation}
\begin{aligned}\langle\hat{M}\rangle & \approx\frac{W_{11}+W_{22}-2\mathrm{Re}W_{12}}{2(1-\mathrm{Re}Y_{12})}+[\frac{y^{2}((W_{11}-\mathrm{Re}W_{12})\cos\theta-\mathrm{Im}W_{12}\sin\theta)+(\mathrm{Re}W_{12}-W_{22})\cos\theta-\mathrm{Im}W_{12}\sin\theta}{y(1-\mathrm{Re}Y_{12})}\\
 & -\frac{((y^{2}-1)(1-\mathrm{Re}Y_{12})\cos\theta-(y^{2}+1)\mathrm{Im}Y_{12}\sin\theta)(W_{11}+W_{22}-2\mathrm{Re}W_{12})}{2y(1-\mathrm{Re}Y_{12})^{2}}]\alpha.
\end{aligned}
\label{eq:55}
\end{equation}

Eq. \eqref{eq:55} shows that the average output of a weak measurement,
$\langle\hat{M}\rangle$, goes linearly as the fidelity $\alpha$
between the pre- and post-selections, when $\alpha\rightarrow0$ and
the pre-selection $|\psi_{i}\rangle\nrightarrow|a_{1}\rangle\,\mathrm{or}\,|a_{2}\rangle.$

Now, let us consider the more complex case: if $|\psi_{i}\rangle\rightarrow|a_{1}\rangle\,\mathrm{or}\,|a_{2}\rangle$
very fast when $\langle\psi_{f}|\psi_{i}\rangle\rightarrow0$, then
the contribution of $x$ must be considered. In this case, how fast
$|\psi_{i}\rangle\rightarrow|a_{1}\rangle\,\mathrm{or}\,|a_{2}\rangle$
is in comparison with $\langle\psi_{f}|\psi_{i}\rangle\rightarrow0$,
plays a critical role in the asymptotic behavior of $\langle\hat{M}\rangle$,
and dominates the competition between $\langle\psi_{f}|\psi_{i}\rangle$
and $x_{1}$ or $x_{2}$.

Suppose $|\psi_{i}\rangle\rightarrow|a_{2}\rangle$, i.e. $y\rightarrow0$.
To characterize the speed of $|\psi_{i}\rangle\rightarrow|a_{2}\rangle$,
we assume that
\begin{equation}
\beta^{s}\sim y,\, s>0,\label{eq:66}
\end{equation}
 as $\alpha\rightarrow0$. The exponent $s$ in \eqref{eq:66} characterizes
the relative speed that the pre-selection $|\psi_{i}\rangle$ tends
towards the eigenstate $|a_{2}\rangle$ of $A$ compared with $\langle\psi_{f}|\psi_{i}\rangle\rightarrow0$,
and determines how fierce the competition between $\langle\psi_{f}|\psi_{i}\rangle$
and $x$ is. It will be shown explicitly below how $s$ affects the
asymptotic behavior of $\langle\hat{M}\rangle$.

Now, by plugging Eq. \eqref{eq:66} into \eqref{eq:23}, one can get
\begin{equation}
\begin{aligned}\langle\hat{M}\rangle & =\frac{\begin{aligned}\beta^{2}(\beta^{4s}W_{11}+W_{22}+2\beta^{2s}\mathrm{Re}W_{12})+\beta^{2s}(W_{11}+W_{22}-2\mathrm{Re}W_{12})\\
+2\beta^{s+1}[\beta^{2s}((W_{11}-\mathrm{Re}W_{12})\cos\theta-\mathrm{Im}W_{12}\sin\theta)+((\mathrm{Re}W_{12}-W_{22})\cos\theta-\mathrm{Im}W_{12}\sin\theta)]
\end{aligned}
}{\begin{aligned}\begin{aligned}\beta^{2}(\beta^{4s}+1+2\beta^{2s}\mathrm{Re}Y_{12})+2\beta^{2s}(1-\mathrm{Re}Y_{12})\\
+2\beta^{s+1}[\beta^{2s}((1-\mathrm{Re}Y_{12})\cos\theta-\mathrm{Im}Y_{12}\sin\theta)+((\mathrm{Re}Y_{12}-1)\cos\theta-\mathrm{Im}Y_{12}\sin\theta)]
\end{aligned}
\end{aligned}
}\end{aligned}
.\label{eq:39}
\end{equation}

Six different powers of $y$ occur in the above equation:
\begin{equation}
\beta^{2+4s},\beta^{2},\beta^{2+2s},\beta^{2s},\beta^{1+3s},\beta^{1+s}.\label{eq:19}
\end{equation}

When $\langle\psi_{f}|\psi_{i}\rangle\rightarrow0$, the two terms
with lowest orders of $\beta$ dominate $\langle\hat{M}\rangle$.
Since the orders of $\beta$ depend on $s$, the asymptotic behavior
of $\langle\hat{M}\rangle$ also depends on $s$, thus in the study
of the asymptotic behavior of $\langle\hat{M}\rangle$ below, the
range of $s$ will be taken into account.

1. $s<1$. For this range of $s$, $\beta^{2s}$ and $\beta^{1+s}$
are the lowest order terms, so
\begin{equation}
\begin{aligned}\langle\hat{M}\rangle & \approx\frac{(W_{11}+W_{22}-2\mathrm{Re}W_{12})/2+\beta^{1-s}((\mathrm{Re}W_{12}-W_{22})\cos\theta-\mathrm{Im}W_{12}\sin\theta)}{\begin{aligned}1-\mathrm{Re}Y_{12}+\beta^{1-s}((\mathrm{Re}Y_{12}-1)\cos\theta-\mathrm{Im}Y_{12}\sin\theta)\end{aligned}
}\\
 & \approx\frac{W_{11}+W_{22}-2\mathrm{Re}W_{12}}{2(1-\mathrm{Re}Y_{12})}+\beta^{1-s}[\frac{(\mathrm{Re}W_{12}-W_{22})\cos\theta-\mathrm{Im}W_{12}\sin\theta}{1-\mathrm{Re}Y_{12}}\\
 & -\frac{((\mathrm{Re}Y_{12}-1)\cos\theta-\mathrm{Im}Y_{12}\sin\theta)(W_{11}+W_{22}-2\mathrm{Re}W_{12})}{2(1-\mathrm{Re}Y_{12})^{2}}],
\end{aligned}
\label{eq:41}
\end{equation}
 and the speed that $\langle\hat{M}\rangle$ varies is
\begin{equation}
\frac{\mathrm{d}\langle\hat{M}\rangle}{\mathrm{d}\beta}\propto\beta^{-s}.\label{eq:77}
\end{equation}

Eq. \eqref{eq:41} shows that the limit of $\langle\hat{M}\rangle$
is \eqref{eq:26}, and this means that in the range $0<s<1$, $y$
only affects the speed that $\langle\hat{M}\rangle$ converges to
\eqref{eq:26} but cannot change the limit of $\langle\hat{M}\rangle$
as $\alpha\rightarrow0$, and the fidelity between the pre-selection
and the post-selection$\langle\psi_{f}|\psi_{i}\rangle$ is still
dominant in its competition with $x$.

2. $s=1$. In this case, $\beta^{2s}$, $\beta^{1+s}$ and $\beta^{2}$
have the same order, so
\begin{equation}
\langle\hat{M}\rangle\approx\frac{W_{11}/2+W_{22}-\mathrm{Re}W_{12}+(\mathrm{Re}W_{12}-W_{22})\cos\theta-\mathrm{Im}W_{12}\sin\theta}{3/2-\mathrm{Re}Y_{12}+(\mathrm{Re}Y_{12}-1)\cos\theta-\mathrm{Im}Y_{12}\sin\theta},\label{eq:20}
\end{equation}
 and
\begin{equation}
\frac{\mathrm{d}\langle\hat{M}\rangle}{\mathrm{d}\beta}\approx0.\label{eq:16}
\end{equation}

So, $\langle\psi_{f}|\psi_{i}\rangle$ and $x$ contribute almost
equally to $\langle\hat{M}\rangle$ as $\alpha\rightarrow0$ when
$s=1$. Intuitively, in this case the contribution from $x$ ``stops''
$\langle\hat{M}\rangle$ running to its limit \eqref{eq:26} as $\alpha\rightarrow0$
and keeps it almost stationary.

3. $s>1$. In this case, $\beta^{1+s}$ and $\beta^{2}$ are the lowest
order terms, so
\begin{equation}
\begin{aligned}\langle\hat{M}\rangle & \approx\frac{W_{22}+2\beta^{s-1}((\mathrm{Re}W_{12}-W_{22})\cos\theta-\mathrm{Im}W_{12}\sin\theta)}{1+2\beta^{s-1}((\mathrm{Re}Y_{12}-1)\cos\theta-\mathrm{Im}Y_{12}\sin\theta)}\\
 & \approx W_{22}+2\beta^{s-1}[((\mathrm{Re}W_{12}-W_{22})\cos\theta-\mathrm{Im}W_{12}\sin\theta)-W_{22}((\mathrm{Re}Y_{12}-1)\cos\theta-\mathrm{Im}Y_{12}\sin\theta)],
\end{aligned}
\label{eq:43}
\end{equation}
 and
\begin{equation}
\frac{\mathrm{d}\langle\hat{M}\rangle}{\mathrm{d}\beta}\propto\beta^{s-2}.\label{eq:29}
\end{equation}

Eq. \eqref{eq:43} shows that the limit of $\langle\hat{M}\rangle$
is $W_{11}$, and this means that in the range $s>1$, the contribution
from $x$ exceeds that from $\langle\psi_{f}|\psi_{i}\rangle$ as
$\alpha\rightarrow0$, and $x$ becomes dominant in its competition
with $\langle\psi_{f}|\psi_{i}\rangle$. Note that $W_{22}$ is the
result of a conventional projective measurement, so in this case,
the weak measurement turns to behave like a conventional projective
measurement in the limit $\alpha\rightarrow0$.

When $\langle\psi_{f}|\psi_{i}\rangle\rightarrow0$ and $s<0$, i.e.
when the pre-selection tends to $|a_{1}\rangle$ as $\alpha\rightarrow0$,
similar results as \eqref{eq:41}-\eqref{eq:43} can be obtained:
\begin{enumerate}
\item $-1<s<0$: only the terms $\beta^{1+3s}$ and $\beta^{2s}$ should
be remained, so
\begin{equation}
\begin{aligned}\langle\hat{M}\rangle & \approx\frac{(W_{11}+W_{22}-2\mathrm{Re}W_{12})/2+\beta^{1+s}((W_{11}-\mathrm{Re}W_{12})\cos\theta-\mathrm{Im}W_{12}\sin\theta)}{1-\mathrm{Re}Y_{12}+\beta^{1+s}((1-\mathrm{Re}Y_{12})\cos\theta-\mathrm{Im}Y_{12}\sin\theta)}\\
 & \approx\frac{W_{11}+W_{22}-2\mathrm{Re}W_{12}}{2(1-\mathrm{Re}Y_{12})}+\beta^{1+s}[\frac{(W_{11}-\mathrm{Re}W_{12})\cos\theta-\mathrm{Im}W_{12}\sin\theta}{1-\mathrm{Re}Y_{12}}\\
 & -\frac{((1-\mathrm{Re}Y_{12})\cos\theta-\mathrm{Im}Y_{12}\sin\theta)(W_{11}+W_{22}-2\mathrm{Re}W_{12})}{2(1-\mathrm{Re}Y_{12})^{2}}],
\end{aligned}
\label{eq:44}
\end{equation}
 and
\begin{equation}
\frac{\mathrm{d}\langle\hat{M}\rangle}{\mathrm{d}\beta}\propto\beta^{s}.\label{eq:40}
\end{equation}

\item $s=-1$: the terms $\beta^{1+3s}$, $\beta^{2+4s}$ and $\beta^{2s}$
should be remained, so
\begin{equation}
\langle\hat{M}\rangle\approx\frac{W_{11}+W_{22}/2-\mathrm{Re}W_{12}+(W_{11}-\mathrm{Re}W_{12})\cos\theta-\mathrm{Im}W_{12}\sin\theta}{3/2-\mathrm{Re}Y_{12}+(1-\mathrm{Re}Y_{12})\cos\theta-\mathrm{Im}Y_{12}\sin\theta},\label{eq:45}
\end{equation}
 and
\begin{equation}
\frac{\mathrm{d}\langle\hat{M}\rangle}{\mathrm{d}\beta}\approx0.\label{eq:42}
\end{equation}

\item $s<-1$: only the terms $\beta^{1+3s}$ and $\beta^{2+4s}$ should
be remained, and
\begin{equation}
\begin{aligned}\langle\hat{M}\rangle & \approx\frac{W_{11}+2\beta^{-(1+s)}((W_{11}-\mathrm{Re}W_{12})\cos\theta-\mathrm{Im}W_{12}\sin\theta)}{1+2\beta^{-(1+s)}((1-\mathrm{Re}Y_{12})\cos\theta-\mathrm{Im}Y_{12}\sin\theta)}\\
 & \approx W_{11}+2\beta^{-(1+s)}[(W_{11}-\mathrm{Re}W_{12})\cos\theta-\mathrm{Im}W_{12}\sin\theta-W_{11}((1-\mathrm{Re}Y_{12})\cos\theta-\mathrm{Im}Y_{12}\sin\theta)],
\end{aligned}
\label{eq:46}
\end{equation}
 and
\begin{equation}
\frac{\mathrm{d}\langle\hat{M}\rangle}{\mathrm{d}\beta}\propto\beta^{-(1+s)}.\label{eq:36}
\end{equation}

\end{enumerate}

Eq. \eqref{eq:41}-\eqref{eq:46} together with \eqref{eq:55} give
the asymptotic value of $\langle\hat{M}\rangle$ when $|\langle\psi_{f}|\psi_{i}\rangle|\rightarrow0$.

\begin{table}[h]
 \centering{}%
\begin{tabular}{|c|c|c|}
\hline
Range of $s$  & $\langle\hat{M}\rangle$  & Convergence Speed of $\langle\hat{M}\rangle$\tabularnewline
\hline
\hline
$0<|s|<1$  & $\beta^{1-|s|}$  & $\beta^{-|s|}$\tabularnewline
\hline
$|s|=1$  & constant  & $0$\tabularnewline
\hline
$|s|>1$  & $\beta^{|s|-1}$  & $\beta^{|s|-2}$\tabularnewline
\hline
$|\psi_{i}\rangle\nrightarrow|a_{1}\rangle$ or $|a_{2}\rangle$  & $\beta$  & $1$\tabularnewline
\hline
\end{tabular}\caption{Asymptotic behavior of $\langle\hat{M}\rangle$ with $\beta^{s}\sim y$
in the limit $\langle\psi_{f}|\psi_{i}\rangle\rightarrow0$.}
\end{table}

The table I summarizes the results of $\langle\hat{M}\rangle$ as
$|\langle\psi_{f}|\psi_{i}\rangle|$ tends towards zero asymptotically,
involving the competition from the limiting process that the pre-selection
approaches one of the eigenstates of the observable $A$. The coefficients
and constant terms are omitted in the table because they do not characterize
the asymptotic behavior of $\langle\hat{M}\rangle$. In contrast to
the prediction by the weak value \eqref{eq:0}, $\langle\hat{M}\rangle$
actually converges as a polynomial of $|\langle\psi_{f}|\psi_{i}\rangle|$
but not the reciprocal of $|\langle\psi_{f}|\psi_{i}\rangle|$ as
$|\langle\psi_{f}|\psi_{i}\rangle|\rightarrow0$.

An interesting thing to note in the table I is that the asymptotic
value of $\langle\hat{M}\rangle$ transits continuously between different
regions of $s$. In the region $0<s<1$ and $-1<s<0$, when $s\rightarrow0$,
$\langle\hat{M}\rangle\rightarrow\beta$ (the coefficients and the
constant term have been omitted), which is exactly the asymptotic
value of $\langle\hat{M}\rangle$ when $|\psi_{i}\rangle\nrightarrow|a_{1}\rangle$
or $|a_{2}\rangle$. In the region $s\neq\pm1$, when $s\rightarrow\pm1$,
$\langle\hat{M}\rangle\rightarrow\mathrm{constant}$, the same as
$s=\pm1$.

\section{Conclusion}

Weak measurement is an interesting quantum measurement scheme with
the ability to amplify tiny physical quantity. The weak value formalism
is valid mainly when the fidelity between the pre- and post-selections
is not too small, and the case that the pre- and post-selections tend
to be orthogonal is beyond the weak value formalism because the weak
value diverges in this case. Our work bridged this gap.

In the first half of this paper, a general rigorous framework for
the weak measurement on a two-dimensional system with a continuous
probe was established, and it was shown that however small the fidelity
between the pre- and post-selections was, the output of a weak measurement
would be always finite. That result set a limit for the amplification
ability of a weak measurement actually. Also two typical examples
were calculated in detail.

In the second half of the paper, the asymptotic behavior of a weak
measurement was considered when the pre- and post-selections tended
to be orthogonal. Generally, the asymptotic behavior of a weak measurement
is dominated by the fidelity between pre- and post-selections which
tends to be zero in the limiting process. But when the pre-selection
tends towards an eigenstate of the observable on the system sufficiently
fast at the same time, the contribution from the fidelity between
the pre-selection and that eigenstate of the observable will be so
prominent that it will enter a competition with the fidelity between
the pre- and post-selections. In the second half of the paper, we
gave a simple model to characterize the speed of the pre-selection
approaching an eigenstate of the observable and analyzed the competition
in detail. The result showed explicitly when the fidelity between
the pre- and post-selections prevailed in the competition and when
the fidelity between the pre-selection and an eigenstate of the observable
did.

We hope this work can contribute to the further understanding and
application of weak measurement.

\section*{Acknowledgement}

This work is supported by the NSFC (Grant No. 11075148 and No. 61125502),
the National Fundamental Research Program (Grant No. 2011CB921300),
the Fundamental Research Funds for the Central Universities, and the
CAS.

\section*{Appendix}

\subsection{Derivation of \eqref{eq:68}}

It is explicit that the real parts of $W_{12}$ and $Y_{12}$ are
\begin{equation}
\begin{aligned}\mathrm{Re}W_{12} & =\int q\mathrm{Re}\phi(q-ga_{1})\mathrm{Re}\phi(q-ga_{2})dq+\int q\mathrm{Im}\phi(q-ga_{1})\mathrm{Im}\phi(q-ga_{2})dq,\\
\mathrm{Re}Y_{12} & =\int\mathrm{Re}\phi(q-ga_{1})\mathrm{Re}\phi(q-ga_{2})dq+\int\mathrm{Im}\phi(q-ga_{1})\mathrm{Im}\phi(q-ga_{2})dq.
\end{aligned}
\label{eq:31}
\end{equation}

Provided that $q\phi^{(n)}(q)\rightarrow0\,(n=0,1,2\cdots)$ as $q\rightarrow\pm\infty$,
it can be straightforwardly verified that
\begin{equation}
\begin{aligned}\int(\mathrm{Re}\phi(q))^{(k)}(\mathrm{Re}\phi(q))^{(2n-k)}dq & =(-1)^{n-k}\int((\mathrm{Re}\phi(q))^{(n)})^{2}dq,\\
\int(\mathrm{Re}\phi(q))^{(k)}(\mathrm{Re}\phi(q))^{(2n+1-k)}dq & =0,\\
\int q(\mathrm{Re}\phi(q))^{(k)}(\mathrm{Re}\phi(q))^{(2n-k)}dq & =(-1)^{n-k}\int q((\mathrm{Re}\phi(q))^{(n)})^{2}dq,\\
\int q(\mathrm{Re}\phi(q))^{(k)}(\mathrm{Re}\phi(q))^{(2n+1-k)}dq & =(-1)^{n-k+1}(n-k+\frac{1}{2})\int((\mathrm{Re}\phi(q))^{(n)})^{2}dq,
\end{aligned}
\label{eq:51}
\end{equation}
 using the method of integration by parts repeatedly. Similar integral
results can be obtained for the imaginary part of $\phi(q)$.

With the Taylor's expansion of $\phi(q-ga_{i})$, i.e.
\begin{equation}
\phi(q-ga_{i})=\sum_{n=0}^{+\infty}\frac{(-1)^{n}}{n!}\phi^{(n)}(q)g^{n}a_{i}^{n},\: i=1,2,\label{eq:32}
\end{equation}
 and the integral formulae \eqref{eq:51}, one can get
\begin{equation}
\begin{aligned}\mathrm{Re}Y_{12} & =\sum_{n=0}^{+\infty}g^{2n}\sum_{k=0}^{2n}\frac{(-1)^{k}(-1)^{2n-k}}{k!(2n-k)!}a_{1}^{k}a_{2}^{2n-k}\int[(\mathrm{Re}\phi(q))^{(k)}(\mathrm{Re}\phi(q))^{(2n-k)}+(\mathrm{Im}\phi(q))^{(k)}(\mathrm{Im}\phi(q))^{(2n-k)}]dq\\
 & =\sum_{n=0}^{+\infty}\frac{(-1)^{n}}{(2n)!}g^{2n}(a_{1}-a_{2})^{2n}\int[((\mathrm{Re}\phi(q))^{(n)})^{2}+((\mathrm{Im}\phi(q))^{(n)})^{2}]dq\\
 & =\sum_{n=0}^{+\infty}\frac{(-1)^{n}}{(2n)!}g^{2n}(a_{1}-a_{2})^{2n}\int|\phi^{(n)}(q)|^{2}dq,\\
\mathrm{Re}W_{12} & =\sum_{n=0}^{+\infty}g^{2n}\sum_{k=0}^{2n}\frac{(-1)^{k}(-1)^{2n-k}}{k!(2n-k)!}a_{1}^{k}a_{2}^{2n-k}\int q[(\mathrm{Re}\phi(q))^{(k)}(\mathrm{Re}\phi(q))^{(2n-k)}+(\mathrm{Im}\phi(q))^{(k)}(\mathrm{Im}\phi(q))^{(2n-k)}]dq\\
 & +\sum_{n=0}^{+\infty}g^{2n+1}\sum_{k=0}^{2n+1}\frac{(-1)^{k}(-1)^{2n+1-k}}{k!(2n+1-k)!}a_{1}^{k}a_{2}^{2n+1-k}\int q[(\mathrm{Re}\phi(q))^{(k)}(\mathrm{Re}\phi(q))^{(2n+1-k)}+(\mathrm{Im}\phi(q))^{(k)}(\mathrm{Im}\phi(q))^{(2n+1-k)}]dq\\
 & =\sum_{n=0}^{+\infty}\frac{(-1)^{n}}{(2n)!}g^{2n}(a_{1}-a_{2})^{2n}\int q|\phi^{(n)}(q)|^{2}dq+\frac{1}{2}\sum_{n=0}^{+\infty}g^{2n+1}\sum_{k=0}^{2n+1}(\frac{(-1)^{n-k}a_{2}}{k!(2n-k)!}a_{1}^{k}a_{2}^{2n-k}-\\
 & \frac{(-1)^{n-k}a_{1}}{(k-1)!(2n+1-k)!}a_{1}^{k-1}a_{2}^{2n-k})\int|\phi^{(n)}(q)|^{2}dq\\
 & =\sum_{n=0}^{+\infty}\frac{(-1)^{n}}{(2n)!}g^{2n}(a_{1}-a_{2})^{2n}\int q|\phi^{(n)}(q)|^{2}dq+\frac{1}{2}\sum_{n=0}^{+\infty}\frac{(-1)^{n}}{(2n)!}g^{2n+1}(a_{1}+a_{2})(a_{1}-a_{2})^{2n}\int|\phi^{(n)}(q)|^{2}dq,
\end{aligned}
\label{eq:52}
\end{equation}
 where we have used the binomial formula
\begin{equation}
(a+b)^{n}=\sum_{k=0}^{n}\frac{n!}{k!(n-k)!}a^{k}b^{n-k}.\label{eq:70}
\end{equation}

\subsection{Derivation of \eqref{eq:69}}

Note that
\begin{equation}
\mathrm{Re}W_{12}=\int\mathrm{Re}\phi(q-ga_{1})\mathrm{Im}\dot{\phi}(q-ga_{2})dq-\int\mathrm{Im}\phi(q-ga_{1})\mathrm{Re}\dot{\phi}(q-ga_{2})dq.\label{eq:37}
\end{equation}
 Provided that $q\phi^{(n)}(q)\rightarrow0\,(n=0,1,2\cdots)$ as $q\rightarrow\pm\infty$,
it can be straightforwardly verified that
\begin{equation}
\begin{aligned}\int(\mathrm{Re}\phi(q))^{(k)}(\mathrm{Im}\phi(q))^{(2n-k)}dq & =(-1)^{n-k}\int(\mathrm{Re}\phi(q))^{(n)}(\mathrm{Im}\phi(q))^{(n)}dq,\\
\int(\mathrm{Re}\phi(q))^{(k)}(\mathrm{Im}\phi(q))^{(2n+1-k)}dq & =(-1)^{n-k}\int(\mathrm{Re}\phi(q))^{(n)}(\mathrm{Im}\phi(q))^{(n+1)}dq,\\
\int(\mathrm{Im}\phi(q))^{(k)}(\mathrm{Re}\phi(q))^{(2n+1-k)}dq & =(-1)^{n-k}\int(\mathrm{Im}\phi(q))^{(n)}(\mathrm{Re}\phi(q))^{(n+1)}dq,
\end{aligned}
\label{eq:30}
\end{equation}
 using the method of integration by parts repeatedly.

Plug the Taylor's expansion of $\phi(q-ga_{i})$ \eqref{eq:32} into
\eqref{eq:37}, it follows that
\begin{equation}
\begin{aligned}\mathrm{Re}W_{12} & =\sum_{n=0}^{+\infty}g^{2n}\sum_{k=0}^{2n}\frac{(-1)^{k}(-1)^{2n-k}a_{1}^{k}a_{2}^{2n-k}}{k!(2n-k)!}(-1)^{n-k}\int[(\mathrm{Re}\phi(q))^{(n)}(\mathrm{Im}\phi(q))^{(n+1)}-(\mathrm{Im}\phi(q))^{(n)}(\mathrm{Re}\phi(q))^{(n+1)}]dq\\
 & +2\sum_{n=0}^{+\infty}g^{2n+1}\sum_{k=0}^{2n+1}\frac{(-1)^{k}(-1)^{2n+1-k}a_{1}^{k}a_{2}^{2n+1-k}}{k!(2n+1-k)!}(-1)^{n+1-k}\int[(\mathrm{Re}\phi(q))^{(n+1)}(\mathrm{Im}\phi(q))^{(n+1)}\\
 & -(\mathrm{Re}\phi(q))^{(n+1)}(\mathrm{Im}\phi(q))^{(n+1)}]dq\\
 & =2\sum_{n=0}^{+\infty}\frac{(-1)^{n}g^{2n}}{(2n)!}(a_{1}-a_{2})^{2n}\int(\mathrm{Re}\phi(q))^{(n)}(\mathrm{Im}\phi(q))^{(n+1)}dq,
\end{aligned}
\label{eq:35}
\end{equation}
 where the binomial formula \eqref{eq:70} has been used.

\end{document}